\documentclass[11pt]{article} 
\usepackage{amsmath}
\usepackage{amsfonts,amssymb,latexsym}

\begin{document}

\newcommand{\nl}{\nonumber\\}
\newcommand{\nnl}{\nl[6mm]}
\newcommand{\nle}{\nl[-2.5mm]\\[-2.5mm]}
\newcommand{\nlb}[1]{\nl[-2.0mm]\label{#1}\\[-2.0mm]}
\newcommand{\ab}{\allowbreak}

\renewcommand{\leq}{\leqslant}              
\renewcommand{\geq}{\geqslant}

\newcommand{\be}{\bes}
\newcommand{\ee}{\ees}
\newcommand{\bes}{\begin{eqnarray}}
\newcommand{\ees}{\end{eqnarray}}
\newcommand{\eens}{\nonumber\end{eqnarray}}

\renewcommand{\/}{\over}
\renewcommand{\d}{\partial}
\newcommand{\e}{{\rm e}}

\newcommand{\mm}{{\mathbf m}}
\newcommand{\kk}{{\mathbf k}}
\newcommand{\xx}{{\mathbf x}}
\newcommand{\yy}{{\mathbf y}}
\newcommand{\qq}{{\mathbf q}}

\newcommand{\jhat}{{\hat\jmath}}

\newcommand{\al}{\alpha}
\newcommand{\bt}{\beta}
\newcommand{\eps}{\epsilon}
\newcommand{\om}{\omega}

\newcommand{\map}{{\mathfrak {map}}}
\newcommand{\g}{{\mathfrak g}}
\newcommand{\G}{{\mathcal G}}
\newcommand{\J}{{\mathcal J}}
\newcommand{\ZZ}{{\mathbb Z}}
\newcommand{\RR}{{\mathbb R}}

\title{{Local, global, divergent -- 
which gauge symmetries are redundant?}}

\author{T. A. Larsson \\
Vanadisv\"agen 29, S-113 23 Stockholm, Sweden\\
email: thomas.larsson@hdd.se}

\maketitle 

\begin{abstract} 
A local gauge symmetry can not possibly be a mere redundancy of the 
description, provided that
\begin{enumerate}
\item
charge is nonzero.
\item
we also consider divergent gauge transformations, whose brackets with
local transformations contain global charge operators.
\end{enumerate}
If these conditions hold, unitarity requires the existence of gauge
anomalies. We describe the relevant multi-dimen\-sional
generalization of affine Kac-Moody algebras, explain why it only
arises if the observer's trajectory is explicitly introduced, and
contrast it to the Mickelsson-Faddeev algebra which pertains to
chiral-fermion type anomalies.
\end{abstract}

\newpage 

Gauge symmetries are widely considered as mere redundancies
of the description rather than as genuine physical symmetries, and
hence all physical states must be gauge invariant. However, this
statement needs qualification. Clearly it can only apply to
local gauge symmetries; global gauge generators include the charge
operators, which act nontrivially on charged states. Thus the claim
is that local gauge symmetries act trivially whereas global
ones do not, at least in the presence of charge.

However, there is a well-known counterexample. The free classical
string has a conformal gauge symmetry, but after quantization it is 
transformed into a conventional, non-gauge symmetry,
which acts nontrivially on the physical Hilbert space, except if
$D=26$, due to conformal anomalies. It is not always appreciated
that also the subcritical ($D<26$) free string is a consistent
theory; according to the no-ghost theorem, the inner product is
positive definite and the action of the Virasoro algebra is 
unitary \cite{GSW86}, despite the nonzero central charge $c=D$.
We are thus led to ask why the gauge symmetries in string theory are 
different from those in other gauge theories.

An infinitesimal gauge transformation (of Yang-Mills type, say) is 
a function valued in a finite-dimen\-sional Lie algebra.
The crucial question is: which class 
of functions? In particular, we need to know the behavior when the
distance from the origin $r \to \infty$. Assume that a function
behaves like $r^n$. We can then distinguish between three classes of
gauge transformations:
\begin{itemize}	
\item
Local, $n < 0$. 
\item
Global, $n = 0$. 
\item
Divergent, $n > 0$.
\end{itemize}
If we only consider local and global transformations, then the local
transformations generate an ideal and the situation is what
we expect: local transformations act trivially, global do not. 
However, if we also consider divergent transformations, the local 
transformations no longer generate an ideal and it is not possible 
to represent them trivially. This is because
\[
[Local, Divergent] = Global + more,
\]
and the RHS is necessarily nonzero for nonzero charge. The best we
can do is to introduce some ground states which are
annihilated by the local subalgebra, and let the divergent
generators be creation operators.

We can also consider gauge transformations with compact support, i.e.
there is some $R < \infty$ such that the functions vanish identically
whenever $r > R$. Informally, we may think that such gauge
transformations behave like $r^{-\infty}$. Hence they annihilate every
state built from finitely many creation operators, so gauge
transformations with compact support do indeed act trivially, even 
when divergent gauge transformations are taken into account. In
contrast, those which vanish at infinity, without being identically
zero for all $r > R$, can not act trivially in the presence of charge.

Let us rephrase this argument in more formal
terms. Consider for definiteness Yang-Mills theory in flat space,
whose gauge symmetries generate the algebra $\G = \map(\RR^3, \dot\g)$
of maps from $3$-dimen\-sional flat space $\RR^3$ to some
finite-dimen\-sional Lie algebra $\dot\g$. $\dot\g$ is defined by
generators $J^a$ and brackets $[J^a, J^b] = if^{ab}{}_c J^c$.
The brackets in $\G$ read
\be
[\J^a(\xx), \J^b(\yy)] = if^{ab}{}_c \J^c(\xx) \delta^3(\xx-\yy),
\ee
where $\xx,\yy \in \RR^3$. Alternatively, we can consider the smeared
generators $\J_X = \int d^3\!x X_a(\xx) \J^a(\xx)$, where $X_a(\xx)$ are
smooth functions on $\RR^3$. $\G$ then takes the form
$[\J_X, \J_Y] = \J_{[X,Y]}$, where 
$[X,Y] = if^{ab}{}_c X_a Y_b J^c$.

$\G$ acts on fields over $\RR^3$ valued in some $\dot\g$ 
representation $R$: 
\be
[\J_X, \psi(\xx)] = X_a(\xx) R(J^a)\psi(\xx).
\ee
$\dot\g$ can be identified with the $\xx$-independent subalgebra 
of $\G$. In particular, the generators $H^i$ of the Cartan 
subalgebra (CSA) constitute a complete set of commuting charges 
$Q^i$: $[H^i, \psi(\xx)] = Q^i \psi(\xx)$. If $\psi(\xx)$ is a charged
field, the CSA, and thus $\dot\g$ itself, clearly acts 
nontrivially on the Hilbert space.
Let $\G_n = \{\J_X \in \G: \exists c < \infty,\hbox{ such that }
|X_a(\xx)| < c|\xx|^n,\ \forall a\hbox{ and }\forall {\xx\in \RR^3} \}$. 
In other words, $\G_n$ consists of gauge transformations which do
not grow faster than $|\xx|^n$ when $\xx \to \infty$.
There is a filtration
\be
\G_{-\infty} \subset ... \subset \G_{-1} \subset \G_0
\subset \G_1 \subset ... \subset \G_\infty = \G,
\ee
where $[\G_m, \G_n] \subset \G_{m+n}$. Hence $\G_{-1}$ is the
subalgebra of local gauge transformations and $\G_0$ the subalgebra
of local+global transformations.
We also consider the associated grading with $\g_n = \G_n/\G_{n-1}$:
\be
\G = ... + \g_{-2} + \g_{-1} + \g_0 + \g_1 + \g_2 + ...,
\label{grading}
\ee
where $[\g_m, \g_n] = \g_{m+n}$. $\g_0$ can clearly be identified
with $\dot \g$. 

The CSA belongs to $\g_0$, and hence $\g_0$ can not act trivially on 
the Hilbert space. Since $[\g_n, \g_{-n}] = \g_0$, the physical Hilbert
space necessarily carries a nontrivial representation of the algebra 
of local gauge transformations in $\G_{-n}$. 
In contrast, if we restrict attention to the local+global algebra
$\G_0$, nothing prevents the local subalgebra $\G_{-1}$ from
acting trivially. 

We can rephrase our argument using the language of induced
representations. Given a $\g_0$ representation $R$, we first extend it
to a representation of the open algebra $\G_0$, also denoted by $R$,
by setting the restriction of $R$ to $\G_{-1}$ to zero. We then
consider the induced $\G$ representation $Ind^{\G}_{\G_0}(R)$. The
restriction of $Ind^{\G}_{\G_0}(R)$ to $\G_{-1}$ is nontrivial even
though the restriction of $R$ is trivial. It is important to note that
$\G$ will in general contain an extension for unitarity.

More explicitly, assume that we can expand all relevant functions in 
a suitable basis, say powers of $r$ times spherical harmonics:
\be
J^a_{n,\ell,m} = r^n Y_{\ell,m}(\theta,\varphi) J^a.
\label{Jar}
\ee
The important issue is the behavior when $r\to\infty$. 
$J^a_{-n,\ell,m}$ are regular when $r\to\infty$, but have a pole
at $r=0$. This pole is irrelevant for the discussion here. It is
straightforward to construct everywhere smooth functions with any
prescribed behavior at infinity, but when we expand functions
in the basis above, we make the additional assumption that the
functions are not just smooth but real-analytic. A real-analytic
function with a zero at infinity necessarily has poles somewhere 
else; hence the spurious poles at the origin.

Since the spherical harmonics form a basis for functions of $\theta$
and $\varphi$, there are coefficients $C^{\ell''}_{\ell,\ell'}$
such that
\be
Y_{\ell,m}(\theta,\varphi) Y_{\ell',m'}(\theta,\varphi) =
\sum_{\ell'' = |\ell-\ell'|}^{\ell+\ell'} C^{\ell''}_{\ell,\ell'}
Y_{\ell'',m+m'}(\theta,\varphi).
\ee
The $J^a_{n,\ell,m}$ satisfy the algebra
\be
[J^a_{n,\ell,m}, J^b_{n',\ell',m'}] = i f^{ab}{}_c 
\sum_{\ell'' = |\ell-\ell'|}^{\ell+\ell'} C^{\ell''}_{\ell,\ell'}
J^c_{n+n',\ell'',m+m'}.
\label{Jalg}
\ee
In particular
\be
[J^a_{n,0,0}, J^b_{-n,0,0}] = i f^{ab}{}_c J^c_{0,0,0}.
\ee
Since $J^a_{0,0,0} = J^a$ can be identified with the generators of the
finite-dimen\-sional algebra $\dot\g$, it acts nontrivially on the
charged Hilbert space. We conclude that both $J^a_{n,0,0}$ and 
$J^b_{-n,0,0}$ also act nontrivially, despite the fact that the
latter is a local gauge transformation which vanishes at infinity.

To avoid the spurious pole at $r=0$, we must employ functions 
that are everywhere smooth and invertible, and still
decay when $r \to \infty$. As an example, consider
\be 
f(r) = 
\begin{cases}
1  &\qquad r \leq 1 \\
1 - \exp(-1/(r-1)) &\qquad r > 1 
\end{cases}
\ee
This function is infinitely differentiable at $r = 1$, since
$\exp(-1/(r-1))$ vanishes faster than any polynomial in $1/(r-1)$
when $r \to 1$. Moreover, $f(r) \approx 1/r$ goes to zero 
when $r \to \infty$, so $J^a_f = f(r) J^a$ generates a local gauge 
transformation. However, because of the non-analyticity at $r=1$,
we cannot expand $J^a_f$ in the basis (\ref{Jar}).
The function $f(r)$ has the inverse
\be 
g(r) = 
\begin{cases}
1  &\qquad r \leq 1 \\
\Big[1 - \exp(-1/(r-1))\Big]^{-1} &\qquad r > 1 
\end{cases}
\ee
The corresponding operator $J^a_g = g(r) J^a$ is
everywhere smooth and well-defined, although it diverges linearly
when $r \to \infty$; it is thus a divergent gauge transformation.
Clearly,
\be
[J^a_f, J^b_g] = if^{ab}{}_c J^c,
\ee
where $J^c$ is the every constant generator of global charge.
Hence $J^a_f$, which generates a local, everywhere smooth gauge
transformation can not be represented trivially if the charge
operator $J^c$ is nonzero.

We note that (\ref{Jalg}) has the subalgebra
\be
[J^a_{n,0,0}, J^b_{n',0,0}] = i f^{ab}{}_c J^c_{n+n',0,0},
\label{KMsub}
\ee
isomorphic to a centerless affine algebra. It is well known that the
only proper unitary lowest-weight representation of this algebra
is the trivial one, for which the charge operator $J^a_{0,0,0} = 0$.
In any unitary representation with nonzero charge, (\ref{KMsub})
acquires a central extension, and hence the full algebra (\ref{Jalg})
is also anomalous\footnote{
This argument is not watertight. If we relax the condition that the
restriction should be of lowest-weight type, there are unitary 
representations with zero central charge, e.g. the direct sum of a
highest- and a lowest-energy representation.}. 
{\em Unitarity and nonzero charge imply anomalies.}

Note the complete analogy with string theory. A basis for a
current algebra on the world sheet is given by $J^a_n = z^n J^a$,
$-\infty < n < \infty$, which classically satisfies the algebra
\be
[J^a_m, J^b_n] = i f^{ab}{}_c J^c_{m+n}.
\label{Jam}
\ee
This algebra is isomorphic to (\ref{KMsub}). In particular, 
(\ref{Jam}) is a classical
gauge symmetry, because a nilpotent BRST operator can always be
constructed classically; anomalies only arise after quantization.
If we additionally require unitarity, it is well known that the charge
generators $J^a_0$ can not be nonzero on the quantum level, unless the
algebra acquires an anomaly, turning it into an affine Kac-Moody
algebra:
\be
[J^a_m, J^b_n] = if^{ab}{}_c J^c_{m+n} 
+ k m \delta^{ab} \delta_{m+n},
\ee
where $\delta^{ab}$ is the Killing metric. By the same token, 
unitarity and nonzero charge also lead to the Kac-Moody-like
extension of the higher-dimen\-sional current algebra, also known
as a toroidal algebra \cite{MEY90,ER05}:
\bes
{[}\J_X, \J_Y] &=& \J_{[X,Y]} + {k\/2\pi i} \delta^{ab}
\int dt\ \dot q^k(t) \d_k X_a(\qq(t)) Y_b(\qq(t)), \nl
{[}\J_X ,q^i(t)] &=& [q^i(t), q^j(t)] = 0.
\label{KM}
\ees

One might argue that transformations with nontrivial behavior at
infinity take us to another superselection sector, which quantum
mechanically is disconnected from the original theory. Since 
different superselection sectors are completely disconnected, we
might as well restrict to a single sector and forget about all the
others. However, this argument begs the question: why is string 
theory different? It is undeniable that in string theory, we consider
gauge generators which diverge at $z=\infty$, both in the Kac-Moody
case (\ref{Jam}) and in the Virasoro case.

A more fruitful viewpoint is that the different superselection sectors
are twisted versions of the original gauge theory. The untwisted
theory is invariant under local gauge transformations, whereas a
divergent transformation takes us to a twisted theory which is not
gauge invariant; instead, there is a local gauge transformation which
takes us back to the untwisted theory. Demanding that not just the 
untwisted theory, but the whole tower of twisted theories over it,
behaves consistently under arbitrary local+global+divergent
gauge transformations is a highly nontrivial constraint, which can 
potentially be used for theory selection.

We can map the situation to a lowest-weight module as follows:

\smallskip
\begin{tabular}{lcl}
untwisted theory &$\Leftrightarrow$& ground state \\
twisted theory &$\Leftrightarrow$& excited state \\
local gauge transformation &$\Leftrightarrow$& lowering operator \\
global gauge transformation &$\Leftrightarrow$& Cartan subalgebra \\
divergent gauge transformation &$\Leftrightarrow$& raising operator
\end{tabular}

\smallskip
\noindent That a gauge symmetry is a redundancy now becomes the
statement that a lowering operator annihilates the ground state.
While true, this is quite uninteresting, because it ignores all 
raising operators and excited states.

An important clarification is necessary. The algebra (\ref{KM})
defines a gauge anomaly in the sense that it is a cohomologically
nontrivial extension of the algebra of gauge transformations.
However, this extension is unrelated to conventional gauge anomalies 
which may arise when chiral fermions are coupled to gauge fields.
The latter type of anomaly is described by a Mickelsson-Faddeev
algebra \cite{Mi89}:
\bes
[\J_X, \J_Y] &=& \J_{[X,Y]} + 
\eps^{ijk} d^{abc} \int d^3\!x\ \d_i X_a(\xx) \d_j Y_b(\xx) A_{ck}(\xx), \nl
{[}\J_X ,A_{ai}(\xx)] &=& if^{bc}{}_a X_b(\xx) A_{ci}(\xx)
+ \d_i X_a(\xx), 
\label{MF}\\
{[}A_{ai}(\xx), A_{bj}(\yy)] &=& 0,
\eens
where $d^{abc}$ are the totally symmetric structure constants
associated with the third Casimir operator. In contrast, the
Kac-Moody-like extension (\ref{KM}) is proportional to the 
{\em second} Casimir $\delta^{ab}$. There is little doubt that
conventional, chiral-fermion gauge anomalies lead to lack of unitarity,
i.e. an inconsistency; a simple argument why the Mickelsson-Faddeev
lacks unitary representations can be found in \cite{Lar05}. This
argument does not apply to the Kac-Moody-like extension (\ref{KM}).

The reason why the Kac-Moody-like anomaly (\ref{KM}) does not arise
in conventional quantization is that the extension is a functional of
a privileged curve $\qq(t) = (q^i(t))$, which can be interpreted as
the trajectory of the observer. Unless such a curve has been
explicitly introduced, it is impossible to formulate the relevant
anomalies, and it becomes impossible to combine unitarity with
nonzero charge within the context of the full algebra of
local+global+divergent gauge transformations.

Let us explain the origin of observer-dependent anomalies,
following \cite{Lar04,Lar05b}.
The basic idea is to first expand all fields 
in a Taylor series around the observer's trajectory, {\em viz.}
\be
\phi(\xx,t) = \sum_{\mm} {1\/\mm!} \phi_{,\mm}(t)(\xx-\qq(t))^\mm,
\label{Taylor}
\ee
where $\mm = (m_1, m_2, m_3)$ 
is a multi-index of length $|\mm| = \sum_{j=1}^3 m_j$, 
$\mm! = m_1!m_2!m_3!$, and $(x-q(t))^\mm =
(x^1-q^1(t))^{m_1} (x^2-q^2(t))^{m_2} (x^3-q^3(t))^{m_3}$.
Up to issues of convergence, it is straightforward to formulate
at least classical physics in terms of Taylor data 
$\{q^i(t), \phi_{,\mm}(t)\}$ rather than field data 
$\phi(\xx,t)$. E.g., the Klein-Gordon equation
\be
{\d^2\phi\/\d t^2} - \nabla^2 \phi + \om^2 \phi = 0
\label{KG}
\ee
translates into the hierarchy (assuming that $\qq(t) = \qq$ is 
time independent)
\be
\ddot\phi_{,\mm}(t) - \sum_{j=1}^3 \phi_{,\mm+2\jhat}(t) 
+ \om^2 \phi_{,\mm}(t) = 0,
\label{KGm}
\ee
where $\jhat$ is a unit vector in the $j$:th direction.
The solution to this hierarchy,
\be
\phi_{,\mm}(t;\kk) = (-i)^{|\mm|} \kk^\mm
\exp(i\sqrt{\om^2+\kk^2}t - i\kk\cdot \qq(t)),
\ee
where $\kk^\mm = k_1^{m_1} k_2^{m_2} k_3^{m_3}$,
corresponds via (\ref{Taylor}) to the field
\be
\phi(\xx,t; \kk) = \exp(i\sqrt{\om^2+\kk^2} t - i\kk\cdot \xx),
\label{plane}
\ee
which of course is a plane-wave solution of (\ref{KG}).
However, the hierarchy (\ref{KGm}) naturally singles out another
class of solutions, of the form 
$\phi(\xx,t) = P(\xx,t)\exp(i\om t)$, $P(\xx,t)$ polynomial, which
are obtained from (\ref{plane}) in the limit $\kk \to 0$. There 
are only finitely many such solutions at any given degree.

In order to quantize a field theory, we must first regularize it.
The natural regularization in a Taylor formulation is to
truncate the Taylor series (\ref{Taylor}) after some finite order $p$,
i.e. we pass to the space of $p$-jets.
However, this comes with a price: the KG hierarchy
(\ref{KGm}) is undefined for $|\mm| = p-1, p$, since it depends on 
the higher-order terms $\phi_{,\mm+2\jhat}(t)$. Therefore, there are
two classes of solutions to the truncated hierarchy:
\begin{itemize}
\item
Finitely many solutions $\phi(\xx,t) = P(\xx,t)\exp(i\om t)$, 
associated with $\mm$ such that $|\mm| \leq p-2$. They exhaust the
plane-waves (\ref{plane}) in the limit $p \to \infty$.
\item
Infinitely many solutions depending on an arbitary choice of 
functions $\phi_{,\mm}(t)$, for $|\mm| = p-1, p$.
\end{itemize}
It is the second class of solutions which is responsible for the 
observer-dependent gauge anomalies (\ref{KM}) in the context of 
gauge theories. This can be regarded as a regularization artefact,
which remains also when the regularization is removed, i.e. in the
limit $p \to \infty$.

The Klein-Gordon equation does of course not possess any gauge
symmetry, but the passage to Taylor data works for gauge theories
as well. Moreover, the truncation to $p$-jets preserves manifest 
gauge and diffeomorphism symmetry classically, since the transformation 
law for a $p$-jet does not involve higher order jets. In fact, it
has been know for a long time that in order to obtain a well-defined
lowest-energy representation of the diffeomorphism algebra, one must
first pass to the space of trajectories in jet space \cite{Lar98}.

There is a simple physical argument why conventional,
observer-dependent quantization runs into trouble when applied to
gravity, in agreement with the known situation.
Namely, in conventional quantum theory there is an implicit
assumption of a classical, macroscopic observer, which is manifest in
the Copenhagen interpretation of quantum mechanics. While such an
assumption is an excellent approximation in most experiments, it
leads to trouble when gravity is taken into account. Namely, a
macroscopic observer is infinitely massive, since she (or at least
her clock, which operationally defines time) does not recoil
as an effect of making an observation. But an infinitely massive
observer will immediately interact with gravity and collapse into a
black hole. To avoid this problem we need an explicitly 
observer-dependent formulation of QFT, along the lines sketched 
above, in order to quantize gravity.

To conclude, we have observed that a local gauge symmetry can not
possibly be a mere redundancy of the description, provided that
\begin{enumerate}
\item
charge is nonzero.
\item
we also consider divergent gauge transformations, whose brackets with
local transformations contain global charge operators.
\end{enumerate}
If these conditions hold, unitarity requires the existence of gauge
anomalies, which in turn requires that the observer's trajectory be
explicitly introduced; this amounts to a modification of QFT.
This seems to be a prerequisite for a meaningful quantum theory 
of gravity.


\begin{thebibliography}{99}

\bibitem{GSW86} M.B. Green, J.H. Schwarz and E. Witten,
   {\it Superstring theory, volume I: Introduction},
   Cambridge Univ. Press (1987).

\bibitem{ER05} S. Eswara Rao,
  {\it On Representations of Toroidal Lie Algebras},
  {\tt math.RT/0503629} (2005)

\bibitem{Lar98} T.A. Larsson,
  {\it Extended diffeomorphism algebras and trajectories in jet space}.
  Comm. Math. Phys. {\bf 214} (2000) 469--491.
  {\tt math-ph/9810003}

\bibitem{Lar05} T.A. Larsson, 
  {\it Why the Mickelsson-Faddeev algebra lacks unitary representations},
  {\tt hep-th/0501023} (2005)

\bibitem{Lar04} T.A. Larsson, 
  {\it Manifestly covariant canonical quantization I: the free 
  scalar field},
  {\tt hep-th/0411028} (2004) 

  --, {\it Manifestly covariant canonical quantization II: Gauge
  theory and anomalies},
  {\tt hep-th/0501043} (2005)

  --, {\it Manifestly covariant canonical quantization III: Gravity,
  locality, and diffeomorphism anomalies in four dimensions},
  {\tt hep-th/0504020} (2006)

  --, in preparation.

\bibitem{Lar05b} T.A. Larsson,        
  {\it Manifestly covariant canonical quantization of gravity and
  diffeomorphism anomalies in four dimensions},
  in {\it Focus on Quantum Gravity Research},
  ed David C. Moore,
  Nova Science Publishers
%  ISBN: 1-59454-660-6 
  (2006).

\bibitem{Mi89} J. Mickelsson,
  {\it Current algebras and groups},
  Plenum Monographs in Nonlinear Physics, London: Plenum Press, 1989.

\bibitem{MEY90} R.V. Moody, S.E. Rao, and T. Yokonoma,
  {\it Toroidal Lie algebras and vertex representations},
  Geom. Ded. {\bf 35} (1990) 283--307.

\end{thebibliography}
\end{document}